\documentstyle[preprint,aps]{revtex}

\let\a=\alpha\let\b=\beta\let\d=\delta
\let\e=\epsilon\let\g=\gamma
\let\l=\lambda

\let\s=\sigma

\newcommand{\be}{\begin{equation}}
\newcommand{\ee}{\end{equation}}
\newcommand{\bea}{\begin{eqnarray}}
\newcommand{\eea}{\end{eqnarray}}

\newcommand{\del}{\partial}

\newcommand{\nbox}{{\,\lower0.9pt\vbox{\hrule \hbox{\vrule height 0.2 cm \hskip
0.2 cm \vrule height 0.2 cm}\hrule}\,}}
\begin{document}
\begin{titlepage}
\begin{center}
\vskip .2in
\hfill
\vbox{
    \halign{#\hfil         \cr
           hep-th/0007241 \cr
           RUNHETC-2000-29 \cr
           }  
      }   
\vskip 0.5cm
{\large \bf Supergravity Solutions for Localised Brane Intersections}\\
\vskip .2in
{\bf Arvind Rajaraman} \footnote{e-mail address:
arvindra@muon.rutgers.edu}
\\
\vskip .25in
{\em
Department of Physics and Astronomy,
Rutgers University,
Piscataway, NJ 08855.  \\}
\vskip 1cm
\end{center}
\begin{abstract}
We present a general method for
constructing supergravity solutions for
intersecting branes. 
The solutions are written in terms of a single function,
which is the solution to a nonlinear differential equation.
We illustrate this 
procedure in detail for
the case of M2-branes ending on M5-branes.
We also present supergravity solutions
for strings ending on Dp-branes.
Unlike previous results in the literature,
these branes are completely localized.
 \end{abstract}
\end{titlepage}
\newpage

\newpage
\bigskip

\section{Introduction}

Supergravity solutions representing D-branes, NS 5-branes, 
and M-branes have been known for a long time now
\cite{Horowitz:1991cd}. It is also known that strings can
end on D-branes
\cite{Polchinski:1995mt}
and branes can end on other branes
\cite{Strominger:1996ac}.
However, supergravity solutions representing such intersecting objects
are, by and large, unknown.

Previous work in the literature has mostly focused on cases
where one of the branes is delocalized
in some direction
\cite{Papadopoulos:1996uq,Youm:1999ti}. There are only a few cases where all the branes
have been localized
\cite{Tseytlin:1996bh,Tseytlin:1996as,Surya:1998dx,Cvetic:2000cj,Hosomichi:2000iz,Itzhaki:1998uz,hash,Gomb,Mar1,Fay1,Fay2},
but these appear to be found on a case
by case basis.

In this paper, we will provide a completely general method for finding
supergravity solutions representing intersecting branes. This
is based on the method introduced by \cite{Fay1}
(See also \cite{Fay2}). 
Using this approach, which we describe in more detail below,
we are able to write the solution in terms of the solution
to a single nonlinear differential equation.
Unfortunately, we have been unable to find
explicit solutions of this  
differential equation. Nevertheless, we regard this as a
solution in principle, and it is far simpler than
attempting to solve the coupled Maxwell-Einstein equations.

The
idea in this approach is to use the supersymmetry equations (the
equations for conserved Killing spinors) as linear equations
determining the field strengths in terms of the metric.
This enables us to write all the fields in terms of
the parameters of the metric.

It is then straightforward to substitute these expressions
into the equations of motion. By the miracle of
supersymmetry, we find that all the equations
of motion simplify into a few simple equations. All the
parameters can then be determined in terms of one
function, which is in turn a solution to a nonlinear
differential equation.
Although we have been unable to find explicit solutions to
this nonlinear equation, nevertheless, we emphasize that
the solution
is completely determined in principle by this differential equation.

We carry out these steps in great detail for the case of M2-branes
ending on M5-branes. 
This method can be easily generalized to all
other systems of intersecting branes in M-theory and string
theory, and as a  further example, we present solutions
for strings ending on Dp-branes.
Finally, we list several open problems.

\section{M2-branes ending on M5-branes}

\subsection{Setup}

The notation will be as follows. Indices with
a tilde over them will denote curved (world)
indices, and indices without a tilde will
be tangent indices.

The M5-brane is oriented along the directions $x_0$ (time), $x_1, x_2, x_3, x_4, x_5$.
The M2-brane is oriented along the directions $x_0, x_1,  x_6$.
Indices $(2, 3, 4, 5)$ will be labeled by
$a, b, \ldots$, and indices $(7, 8, 9, 10)$ will be
labeled $\a, \b, \ldots$. 

The M5-brane and M2-brane act as sources for $a^{(3)}_{\a\b\g}$
and $a^{(3)}_{016}$ respectively. Correspondingly, we expect the field strengths
$f_{\a\b\g\d},f_{\a\b\g a}, f_{\a\b\g 6}$, and 
$f_{01a6},  f_{016\a}$ to be nonzero. Furthermore, the
M2 brane acts as a source for the $U(1)$ field on the
worldvolume of the M5-brane. If the worldvolume field strength
is denoted by $h$, then $h_{01a}$ and $h_{abc}$ are nonzero.
This field acts as a source for  $a^{(3)}_{abc}$ and
 $a^{(3)}_{01a}$. Correspondingly, we also expect
$f_{01a\a}$, and $f_{2345}, f_{abc6}, f_{abc\a}$
to be nonzero.

Also, since $\del_a X^6$ is nonzero on the M5 brane
worldvolume, we expect $g_{a6}$ to be nonzero.

The nonzero components of the field strengths
will be denoted
\bea
F_{\a\b\g\d},\quad F_{\a\b\g a},\quad F_{\a\b\g 6};
\quad
G_{01a6}, \quad G_{01a\a},\quad G_{016\a};
\quad
H_{2345},\quad H_{abc6},\quad H_{abc\a}
\nonumber
\eea
We have chosen a notation where different components of
the field strength are denoted by $F, G,$ or $H$ depending
on the indices.

The symmetry
further dictates that $(U=\sqrt{x_\a^2})$
\bea
F_{8910a}={x_7\over U}F_a
\quad
F_{7910a}=-{x_8\over U}F_a
\quad
F_{7810a}={x_9\over U}F_a
\quad
F_{789a}=-{x_{10}\over U}F_a
\nonumber
\\
F_{89106}={x_7\over U}F_6
\quad
F_{79106}=-{x_8\over U}F_6
\quad
F_{78106}={x_9\over U}F_6
\quad
F_{7896}=-{x_{10}\over U}F_6
\\
H_{abc\a}={x_\a\over U}H_{abc}
\quad
G_{016\a}={x_\a\over U}G_6
\quad
G_{01a\a}={x_\a\over U}G_a
\nonumber
\eea

\subsection{The Equations for Supersymmetry}

We can now analyse the equation for
conserved supercharges
\bea
\d \psi_{\tilde{\mu}}= \del_{\tilde{\mu}} \e - {1 \over 4}\omega_{\tilde{\mu}}^{ab}\gamma_{ab}
+{i\over 288}(\gamma_{\tilde{\mu}}^{\a\b\g\d}-8\d_{\tilde{\mu}}^\a\gamma^{\b\g\d})
f_{\a\b\g\d}\e=0
\eea

These equations are rather cumbersome.
For example, for $\mu=1$, we have
\bea
\left({1 \over 2}\omega_1^{1a}\gamma_a+{1 \over 2}\omega_1^{16}\gamma_6
+{1 \over 2}\omega_1^{1\a}\gamma_\a\right)\e
-{i \over 12}F_{78910}\gamma^{78910}\e
-{i \over 12}H_{2345}\gamma^{2345}\e~~~~~~~~~~~~~~
\nonumber
\\
-{i \over 12}\left(F_{789a}\gamma^{789a}+F_{7810a}\gamma^{7810a}
+F_{7910a}\gamma^{7910a}+F_{8910a}\gamma^{8910a}\right)\e~~~~~~~~~~~~~
\nonumber
\\
-{i \over 12}\left(F_{7896}\gamma^{7896}+F_{78106}\gamma^{78106}
+F_{79106}\gamma^{79106}+F_{89106}\gamma^{89106}\right)\e~~~~~~~~~~~~~~
\nonumber
\\
-{i \over 12}\left(H_{234\a}\gamma^{234\a}+H_{235\a}\gamma^{235\a}
+H_{245\a}\gamma^{245\a}+H_{345\a}\gamma^{345\a}\right)\e~~~~~~~~~~~~~~
\nonumber
\\
-{i \over 12}\left(H_{2346}\gamma^{2346}+H_{2356}\gamma^{2356}+
H_{2456}\gamma^{2456}+H_{3456}\gamma^{3456}\right)\e~~~~~~~~~~~~~~
\nonumber
\\
+{i \over 6} \left(G_{01a\a}\gamma^{01a\a}+G_{01a6}\gamma^{01a6}+
G_{016\a}\gamma^{016\a}\right) \e=0
\eea
All indices above are tangent space indices; we have defined
$\omega_a^{bc}=e_a^{\tilde{\mu}}\omega_{\tilde{\mu}}^{bc}$.

To simplify the problem, we note that since
we are looking for BPS solutions, we expect to be able
to superpose them. In this case, we should be able to
superpose M2 branes with arbitrary coordinates in
the directions $x_a$. 
We can therefore separate the SUSY equations into terms which
are even in $x_a$, and terms which are odd in $x_a$.
These equations should be separately satisfied.
The terms involving a $\del_a$ clearly produces a term 
odd in $x_a$, while $\del_6, \del_\a$ produce terms 
even in $x_a$.

In addition, we will take the ansatz for
the spinor
to be
\bea
\e=(g_{11})^{1\over 4}\e_0
\eea
where $\e_0$ is a constant spinor. This ansatz is
justified in 
\cite{Kallosh:1997qw}.

The parts of the equations odd in $x_4$ are
\bea
\label{susyodd}
{1 \over 2}\omega_1^{14}\gamma_4\e+\left(-{i \over 12}H_{235\a}\gamma^{235\a}
-{i \over 12}H_{2356}\gamma^{2356}+{i \over 6}G_{014\a}\gamma^{014\a}
+{i \over 6}G_{0146}\gamma^{0146}\right)\e~~~~~~~~~~~~~~ \nonumber
\\
-{i \over 12}\left(F_{7894}\gamma^{7894}+F_{78104}\gamma^{78104}
+F_{79104}\gamma^{79104}+F_{89104}\gamma^{89104}\right)\e=0
\nonumber
\\
{1 \over 2}(\omega_5^{54}-\omega_1^{14})\gamma_4\e
+\left({i \over 4}H_{235\a}\gamma^{235\a}
+{i \over 4}H_{2356}\gamma^{2356}
-{i \over 4}G_{014\a}\gamma^{014\a}
-{i \over 4}G_{0146}\gamma^{0146} \right)\e=0
\nonumber
\\
{1 \over 2}\left(\omega_5^{54}\gamma_4+\omega_4^{67}\gamma_{467}\e\right)+
\left({i \over 4}H_{235\a}\gamma^{235\a}
+{i \over 4}H_{2356}\gamma^{2356}
\right)\e
~~~~~~~~~~~~~~~~~~~~~~~~~~~~~~~~~~~~
\nonumber
\\
\left(-{i \over 4}G_{014\a}\gamma^{014\a}
-{i \over 4}G_{0146}\gamma^{0146}
-{i \over 4}F_{89104}\gamma^{89104}\right)\e=0
\\
{1 \over 2}(\omega_6^{64}-\omega_1^{14})\gamma_4\e
-{1 \over 2}\omega_6^{4\a}\gamma_{64\a}\e
+\left({i \over 4}H_{2356}\gamma^{2356}
-{i \over 4}G_{014\a}\gamma^{014\a}\right) \e=0
\nonumber
\\
-{1 \over 2}\omega_7^{64}\gamma_{764}
+\left({i \over 4}G_{0147}\gamma^{0147}
+{i \over 4}H_{2357}\gamma^{2357}
-{i \over 4}F_{89104}\gamma^{89104}\right) \e=0
\nonumber
\eea

The parts of the equations even in all $x_a$ are
\bea
\label{susyeven}
{1 \over 2}\omega_1^{16}\gamma_6\e
+{1 \over 2}\omega_1^{1\a}\gamma_\a\e
+\left({i \over 6}G_{016\a}\gamma^{016\a}
-{i \over 12}H_{2345}\gamma^{2345}
-{i \over 12}F_{78910}\gamma^{78910}\right)\e
~~~~~~~~~~~~~~~~~~~~~	
\nonumber
\\
-{i \over 12}\left(F_{7896}\gamma^{7896}+F_{78106}\gamma^{78106}
+F_{79106}\gamma^{79106}+F_{89106}\gamma^{89106}\right)\e=0
\nonumber
\\
{1 \over 2}(\omega_5^{56}-\omega_1^{16})\gamma_6\e
+{1 \over 2}(\omega_5^{5\a}-\omega_1^{1\a})\gamma_\a\e
+\left({i \over 4}H_{2345}\gamma^{2345}-{i \over 6}G_{016\a}\gamma^{016\a}\right)\e=0
\nonumber
\\
{1 \over 2}(\omega_6^{6\a}-\omega_1^{1\a})\gamma_\a\e
+{i \over 4}\left(F_{7896}\gamma^{7896}+F_{78106}\gamma^{78106}
+F_{79106}\gamma^{79106}+F_{89106}\gamma^{89106}\right)\e=0
\\
{1 \over 2}(\omega_1^{17}-\omega_8^{87})\gamma_7\e
+{i \over 4}G_{0167}\gamma^{0167}\e
-{i \over 4}F_{89106}\gamma^{89106}\e=0
\nonumber
\eea

These equations should preserve one quarter of the
supersymmetries. In other words, each equation should be
proportional to a linear combination of $P_1\e_0$ and
$P_2\e_0$, where $P_1\e_0=P_2\e_0=0$ and $P_1, P_2$ are
projection operators.

A glance at the equations (\ref{susyodd}) and (\ref{susyeven}) shows that the
only possible choice of projection operators is (upto signs)
\bea
\label{projs}
P_1\e_0\equiv (1+ i\g^{678910})\e_0=0
\nonumber
\\
P_2\e_0\equiv (1+ i\g^{016})\e_0=0
\eea
With these projection operators, we can reduce the
matrix equations (\ref{susyodd}) and (\ref{susyeven}) to a set of algebraic
equations.
We will treat these algebraic equations as
equations determinining the various field strengths
as functions of the spin connections (and thereby
implicitly as functions of the metric.)

The equations for the field strengths are then
\bea
G_{0164}=4w_1^{14}+2w_5^{54}
\\
G_{0167}=2w_1^{17}-2w_4^{47}
\\
G_{0147}=2w_4^{67}
\\
H_{2356}=4w_5^{54}+2w_1^{14}
\\
H_{2357}=-2w_4^{67}
\\
H_{2345}=2w_{4}^{46}-2w_1^{16}
\\
F_{48910}=-2w_4^{67}
\\
F_{78910}=-2w_{4}^{46}-w_1^{16}
\\
F_{68910}=4w_4^{47}+2w_1^{17}
\eea

In addition, we get the constraints
\bea
w_1^{16}+w_{4}^{46}+w_7^{76}=0
\\
w_1^{18}+w_{4}^{48}+w_7^{78}=0
\\
w_1^{14}+w_{5}^{54}+w_7^{74}=0
\\
w_4^{67}=w_6^{47}=-w_7^{64}
\\
w_6^{67}+2w_4^{47}=0
\\
w_6^{64}=2w_{5}^{54}+2w_1^{14}
\eea

In all the equations above, all the indices are tangent space
indices; we have defined $w_c^{ab}=e_c^{\tilde{\mu}} w_{\tilde{\mu}}^{ab}$.

We can solve all the constraints above by the metric ansatz
\bea
\label{metric}
e_{0\tilde{0}}=e_{1\tilde{1}}=\l^{-{1\over 3}}H^{-{1\over 6}}
\nonumber
\\
e_{2\tilde{2}}=e_{3\tilde{3}}=e_{4\tilde{4}}=e_{5\tilde{5}}=\l^{1\over 6}H^{-{1\over 6}}
\nonumber
\\
e_{6\tilde{6}}=\l^{-{1\over 3}}H^{1\over 3}
\\
e_{7\tilde{7}}=e_{8\tilde{8}}=e_{9\tilde{9}}=e_{10\tilde{10}}=\l^{1\over 6}H^{1\over 3}
\nonumber
\\
e_{6\tilde{a}}=\phi_a e_{6\tilde{6}}
\nonumber
\eea
with the constraint
\bea
\label{eom0}
\del_6(H\phi_a)=\del_aH
\eea

\subsection{The Equations of Motion}

We now look at the equations of motion. Away from
the M5-brane, we can look at the vacuum equations of motion,
which for the gauge fields is
\bea
\del_\mu f^{\mu\nu\rho\s}={1\over 2.(24)^2}\e^{\nu\rho\s 
\tilde{a}\tilde{b}\tilde{c}\tilde{d}\tilde{e}\tilde{f}\tilde{g}\tilde{h}}
f_{\tilde{a}\tilde{b}\tilde{c}\tilde{d}}f_{\tilde{e}\tilde{f}\tilde{g}\tilde{h}}
\eea

We can now substitute the field strengths found above
into these equations. Remarkably, all the equations of motion
collapse to the equations
\bea
\label{eom1}
\del_a\left( \phi_a\right)
-{1\over H}\del_6\l-{1\over 2}\del_6 \phi_a^2 =0
\eea
and
\bea
\del_\a^2\left(H\right)+\del_6^2(H\l)=0
\eea

When we include sources, we need to modify the second equation. 
Defining
\bea
\label{eom2}
\del_\a^2\left(H\right)+\del_6^2(H\l)=\del_6 Q
\eea
we have
\bea
\del_7F_{a8910}-\del_8F_{a7910}+\del_9F_{a7810}-\del_{10}F_{a789}
-\del_a F_{78910}= \del_a Q
\\
\del_\mu G^{\mu 01a}+  
{1\over 2.(24)^2}\e^{\tilde{0}\tilde{1}\tilde{a}
\tilde{A}\tilde{B}\tilde{C}\tilde{D}\tilde{E}\tilde{F}\tilde{G}\tilde{H}}
f_{\tilde{A}\tilde{B}\tilde{C}\tilde{D}}
f_{\tilde{E}\tilde{F}\tilde{G}\tilde{H}}= \del_a Q -i\phi\del_6 Q
\\
\del_6\left( \del_\mu G^{\mu 016}+
{1\over 2.(24)^2}\e^{\tilde{0}\tilde{1}\tilde{6}
\tilde{A}\tilde{B}\tilde{C}\tilde{D}\tilde{E}\tilde{F}\tilde{G}\tilde{H}}
f_{\tilde{A}\tilde{B}\tilde{C}\tilde{D}}
f_{\tilde{E}\tilde{F}\tilde{G}\tilde{H}}\right)=-\del_a^2Q
\eea

$Q$ thus parametrizes the perturbation of the original 
M5-brane due to the addition of the M2-brane.

\subsection{Linearized analysis}

At the linearized level, we can drop the terms involving
$\phi$.
If we then take
\bea
\del_\mu G^{\mu \tilde{0}\tilde{1}\tilde{a}}+
{1\over 2.(24)^2}\e^{\tilde{0}\tilde{1}\tilde{a}
\tilde{A}\tilde{B}\tilde{C}\tilde{D}\tilde{E}\tilde{F}\tilde{G}\tilde{H}}
f_{\tilde{A}\tilde{B}\tilde{C}\tilde{D}}
f_{\tilde{E}\tilde{F}\tilde{G}\tilde{H}}
 = \delta(x_\a)\d(x_6)h^{01a}
\\
\del_\mu G^{\mu \tilde{0}\tilde{1}\tilde{6}}+
{1\over 2.(24)^2}\e^{\tilde{0}\tilde{1}\tilde{6}
\tilde{A}\tilde{B}\tilde{C}\tilde{D}\tilde{E}\tilde{F}\tilde{G}\tilde{H}}
f_{\tilde{A}\tilde{B}\tilde{C}\tilde{D}}
f_{\tilde{E}\tilde{F}\tilde{G}\tilde{H}}
=\cases{\d^4(x_a)\d^4(x_\a), & $x_6\ge 0$;\cr
0, & else.\cr}
\eea
the above equations are
satisfied provided
\bea
\del_ah^{01a}=\d(x_a)
\eea

In other words, we have a gauge field strength on the
M5-brane corresponding to the field of an electric charge
on a line at $x_a=0$. Also, we have a membrane
(the source for $A_{016}$) extending along the 
positive $x_6$ axis. This is exactly the configuration of a M2-brane
ending on an M5-brane.

At the linearized level, the equations (\ref{eom0}), (\ref{eom1})
and (\ref{eom2}) also
simplify to
\bea
\del_6 (H\phi_a)=\del_aH
\\
\del_a(H\phi_a)=\del_6\l
\\
\del_\a^2(H\phi_a)+\del_a\del_6(H\l)=\del_aQ
\eea
which can be simplified to the linear equation
\bea
H\del_a^2\l+\del_6^2\l+\del_\a^2\l={1\over \del_6} \del_a^2Q
=\cases{\d^4(x_a)\d^4(x_\a), & $x_6\ge 0$;\cr
0, & else.\cr}
\eea

More generally, we have multiple M2 branes ending
at different points on the M5-brane. The general
solution,
at the linear level, is then
\bea
\label{linsol}
\l=\int dx' K(x,x') \rho(x')
\eea
where we have introduced the Green's function K, satisfying
\bea
(H_0\del_a^2+\del_6^2+\del_\a^2)K(x,x')=\d(x-x')
\eea
$H_0$ is the harmonic function describing the
M5-branes without the M2-branes 
and $\rho(x')$ is the membrane source density.

The full nonlinear solution no longer has pointlike
sources. The source $Q$ is now a nontrivial function of
$x_a$ and $x_6$, and even the identification of $Q$ as
a worldvolume field is a little problematic.

However, at long distances from the M2-branes, the linearized
analysis still applies, since $\l$ is small. Also,
very close to any M2-brane, where $\l$ diverges, the
solution (\ref{linsol}) should still apply, since the effects of the
membrane dominate.

At the nonlinear level, we can then say that $\l$ is
a solution to the equations  (\ref{eom0}), (\ref{eom1})
and (\ref{eom2}), subject to the
condition that 
\bea
\l=\int dx' K(x,x') \rho(x')
\eea
both when $\l$ is small or when it is very large.
This is expected to completely specify
the solution, although we cannot explicitly solve
the equations.

\subsection{Comments}

1. One might ask why one needs to look at the equations
of motions at all. After all, once we impose the 
constraint (\ref{eom0}), we have a solution that is
supersymmetric. It is well known that a solution
preserving a supersymmetry should also satisfy the
equations of motion.

The point is that imposing the supersymmetry constraints
does not completely specify the sources. The M2-branes 
along $x_0, x_1, x_6$, and the M5-branes along 
$x_0, x_1, x_2, x_3, x_4, x_5,$ preserve the 
supersymmetries (\ref{projs}). But one can
also add M5-branes oriented along $x_0, x_1,
x_7, x_8, x_9, x_{10}$ without breaking any
more supersymmetries. The equation (\ref{eom1})
precisely sets the number of these M5-branes to
zero, so that we have the sources we started with.

2. The second equation (\ref{eom2}) determines the
positions of the M5-brane and M2-brane sources.

One expects on physical grounds that the M5-branes
should not deform in the $x_\a$ directions. In this 
case, one should find that $Q=\d^4(x_\a) Q_1(x_a, x_6)$.

It is however not clear how to show that the differential
equations (\ref{eom0}), (\ref{eom1}), (\ref{eom2}) are consistent
with this expectation. One could, in principle, worry that
when one tried to integrate in the equations from $x_\a=\infty$,
one gets a singularity away from the origin. The solution
would then be valid only outside this region (somewhat
like the enhancon phenomenon 
\cite{Johnson:2000qt}).
It would be extremely surprising (at least to this author)
if such a phenomenon should occur in this system, so
we shall assume that in fact, $Q=\d^4(x_\a) Q_1(x_a, x_6)$.

\subsection{Summary}
The solution for M2-branes ending on M5-branes
is given by the metric (\ref{metric})
\bea
ds^2=\l^{-{2\over 3}}H^{-{1\over 3}}(-dx_0^2+dx_1^2)
+\l^{1\over 3}H^{-{1\over 3}}(dx_2^2+dx_3^2+dx_4^2+dx_5^2)
\nonumber
\\
+\l^{1\over 3}H^{2\over 3}(dx_7^2+dx_8^2+dx_9^2+dx_{10}^2)
+\l^{-{2\over 3}}H^{2\over 3}(dx_6+\phi_adx_a)^2
\eea
The field strengths are given in equations (8)-(16).

We also have the equations (\ref{eom0}), (\ref{eom1}), 
and
(\ref{eom2}) which determine the parameters
\bea
\del_6(H\phi_a)=\del_aH
\\
\del_a\left( \phi_a\right)
-{1\over H}\del_6\l-{1\over 2}\del_6 \phi_a^2 =0
\\
\del_\a^2\left(H\right)+\del_6^2(H\l)=\d^4(x_\a)\del_6 Q
\eea
where $Q$ is a function of $x_a, x_6$.

Far away from the M2-branes, we have the boundary condition
\bea
H\rightarrow H_0\qquad \phi_a\rightarrow 0\qquad \l\rightarrow 1
\eea
The first linear perturbation in $\l$ is
\bea
\label{lbeh}
\l=\int dx' K(x,x') \rho(x') +1
\eea
where $ \rho(x')$ is the membrane source density, and $K(x,x')$
is the Green's function satisfying
\bea
(H_0\del_a^2+\del_6^2+\del_\a^2)K(x,x')=\d(x-x')
\eea
where $H_0$ is the harmonic function describing the
M5-brane background without the M2-branes.
Furthermore, $\l$ is expected to behave as the above equation (\ref{lbeh})
in regions where $\l$ diverges.

The system of equations 
(\ref{eom0}), (\ref{eom1}),
and
(\ref{eom2}) can further be simplified to the following equations
\bea
H\phi_a=\del_a\del_6\tau\\
H=\del_6^2\tau\\
\l=\del_a^2\tau-H\phi_a^2
\eea
where the function $\tau$ satisfies the nonlinear differential
equation
\bea
\del_\a^2\tau +\del_6^2\tau\del_a^2\tau-(\del_a\del_6\tau)^2
=\d^4(x_\a){1\over \del_6} Q
\eea

\section{Strings ending on Dp-branes}

We can carry out a very similar analysis for
strings ending on Dp-branes. We shall take
the Dp-branes oriented along the directions
$x_0..x_p$, the string is oriented along
$x_0,x_9$.

The Killing spinors preserved by this configuration are
given by $\e=(g_{00}^{1/4})\e_0$, where $\e_0$ is a constant
spinor satisfying
\bea
\gamma^{09}\e_0=\eta_1\e_0
\\
\gamma^{01..p}\e_0=\eta_2\e_0
\eea
where $\eta_1, \eta_2$ are constants.

The Killing equations can be written in the generic form
\bea
w_\mu^{ab}\gamma_{ab}\e+(a_GG_{\mu ab}\gamma^{ab}+b_GG_{abc}\gamma_\mu^{abc})\e
+
(a_HH_{\mu a_1..a_{p-2}}\gamma^{a_1..a_{p-2}}+b_HH_{a_1..a_{p-1}}\gamma_\mu^{a_1..a_{p-1}})\e
\nonumber
\\
+(a_FF_{\mu a_1..a_{p}}\gamma^{a_1..a_{p}}+b_FF_{a_1..a_{p}}\gamma_\mu^{a_1..a_{p}})\e
=0
\eea

The metric components are found by noting that
the diagonal components are the products of the
corresponding component in the metric of the string and the
p-brane metric (i.e. $g_{ii}=g_{ii}^{(string)}g_{ii}^{(p-brane)}$.)
(We will work in the string frame)
\bea
e_{0\tilde{0}}=\l^{-{1\over 2}}H^{-{1\over 4}}
\nonumber
\\
e_{1\tilde{1}}=e_{2\tilde{2}}=..=e_{p\tilde{p}}=H^{-{1\over 4}}
\nonumber
\\
e_{9\tilde{9}}=\l^{-{1\over 2}}H^{1\over 4}
\\
e_{(p+1)\tilde{(p+1)}}=..=e_{8\tilde{8}}=H^{1\over 4}
\nonumber
\eea
In addition, we have an off diagonal
component.
\bea
e_{9\tilde{a}}=\phi_a e_{9\tilde{9}}
\nonumber
\eea

The dilaton is
\bea
e^\phi=\l^{-1/2}H^{-{p-3\over 4}}
\eea

The equations for the field strengths are then
(all indices are tangent space indices)
\bea
(a_G\eta_1)G_{019}=-w_2^{21}-w_0^{01}
\\
(a_G\eta_1)G_{098}=w_0^{08}-w_2^{28}
\\
(a_G\eta_1)G_{018}=w_8^{19}
\\
(a_H\eta_1\eta_2)e^\phi H_{92..p}=-w_2^{21}
\\
(a_H\eta_1\eta_2)e^\phi H_{82..p}={1\over 2}w_8^{19}
\\
(a_H\eta_1\eta_2)e^\phi H_{1..p}={1\over 2}(w_2^{29}-w_0^{09})
\\
(a_F\eta_2)e^\phi F_{801..p}=w_2^{28}
\\
(a_F\eta_2)e^\phi F_{8902..p}=-w_8^{19}
\\
(a_F\eta_2)e^\phi F_{901..p}={1\over 2}(w_0^{09}+w_2^{29})
\eea

The equations relating the various parameters are 
\bea
\del_9(H\phi_a)=\del_aH
\\
H^{-1}\del_9\l+{1\over 2}\del_9(\phi_a^2)=\del_a\phi_a
\\
\del_9^2(H\l)+\del_\a^2(H)=\d(x_\a)\del_9Q
\eea
which can be simplified in terms of a single function $\tau$
\bea
H=\del_9^2\tau
\\
H\phi_a=\del_a\del_9\tau
\\
\l=\del_a^2\tau-H\phi_a^2
\eea
where $\tau$ satisfies
\bea
\del_\a^2\tau+\del_9^2\tau\del_a^2\tau-(\del_a\del_6\tau)^2=\d(x_\a)
{1\over \del_9}Q
\eea
which is the same differential equation as in the previous section.

\section{Open questions}

We have reduced the construction of intersecting
brane systems to solving a single nonlinear
partial differential equation. Much of the
physics is contained in the solution to
this equation, which, unfortunately, we have been unable to
find exactly. It would be useful
to be able to extract some information from the
equation, for instance, to know whether it admits
multiple solutions in some cases.

There are several potential applications of these solutions.
By the AdS/CFT correspondence \cite{Maldacena:1998re}, some of these solutions
can be mapped to configurations in strongly coupled gauge
theories. For example, the solution for $k$ D-strings ending on $N$ D3-brane
is dual to a configuration of $k$ monopoles in
$SU(N)$ ${\cal N}=4$ Yang-Mills theory. 
We can then extract quantities like the quark
monopole potential etc. in this limit.

Similarly, these configurations, can be used
to construct supergravity duals for any gauge
theory that can be engineered by putting branes
on orbifolds. These include pure ${\cal N}=2$ gauge theories
and ${\cal N}=1$ gauge theories. One can also study solitons
in these theories.

One very interesting open question is to find the
near extremal solutions. It is, of course, not
possible to use the supersymmetry equations there,
but the solutions found here should be a useful
starting point to get the nonextremal solutions.
Such solutions would give us information about
the thermodynamics of theories with less supersymmetry.

\section{Acknowledgements}

This research was supported in part by DOE grant DE-FG02-96ER40559.

\end{document}